\documentclass[runningheads]{llncs}
\usepackage{hyperref}
\usepackage{algorithm2e}
\usepackage{graphicx}
\usepackage{caption}
\usepackage{subcaption}
\usepackage[utf8]{inputenc}
\begin{document}
\title{A Blockchain-based Decentralized Self-balancing Architecture for the Web of Things
}

%
%
\author{Aleksandar Tošić\inst{1,2}
\and
Jernej Vičič\inst{2}
\and
Michael Mrissa\inst{1,2}
}
\authorrunning{A. Tošić et al.}
%
\institute{
InnoRenew CoE,\\
Livade 6, 6310 Izola, Slovenia\\
\email{\{firstname.surname\}@innorenew.eu}
\and
University of Primorska,\\
Faculty of Mathematics, Natural Sciences and Information Technology,\\
Glagoljaška ulica 8, 6000 Koper, Slovenia\\
\email{\{firstname.surname\}@famnit.upr.si}
}
\maketitle              
\begin{abstract}
Edge computing is a distributed computing paradigm that relies on computational resources of end devices in a network to bring benefits such as low bandwidth utilization, responsiveness, scalability and privacy preservation.
Applications range from large scale sensor networks to IoT, and concern multiple domains (agriculture, supply chain, medicine\dots).
However, resource usage optimization, a challenge due to the limited capacity of edge devices, is typically handled in a centralized way, which remains an important limitation.
In this paper, we propose a decentralized approach that relies on a combination of blockchain and consensus algorithm to monitor network resources and if necessary, migrate applications at run-time.
We integrate our solution into an application container platform, thus  providing an edge architecture capable of general purpose computation.
We validate and evaluate our solution with a proof-of-concept implementation in a national cultural heritage building.

\end{abstract}
\section{Introduction}

In the last few years, edge computing has received a lot of attention as an alternative to cloud computing, due to the multiple advantages it offers, such as low bandwidth usage, responsiveness, scalability~\cite{mach2017mobile} and privacy preservation~\cite{satyanarayanan2017emergence}.
Edge computing now becomes possible due to the evolution of devices that offer more computational power than ever.
Combined with application container platforms such as Docker~\cite{anderson2015} that mask heterogeneity problems, it becomes possible for connected devices to form a homogeneous distributed run-time environment.
Additionally, orchestration engines (i.e. Kubernetes\footnote{\url{https://kubernetes.io/}}) have been developed to manage and optimize usage of network, memory, storage or processing power for edge devices and improve the global efficiency, scalability and energy management of edge platforms.
However, such solutions are centralized, which means that they represent a single point of failure (SPOF), which entails several drawbacks, such as lack of reliability and security.
The problem is so critical that developments for high availability have been explored, for instance with Kubernetes\footnote{\url{https://kubernetes.io/docs/setup/independent/setup-ha-etcd-with-kubeadm}}.

In this paper, we propose to tackle this problem with a decentralized algorithm that monitors network resources to drive application execution.
Our solution relies on an original combination of blockchain-like shared data structure, consensus algorithm and containerized monitoring application to enable run-time migration of applications, when relevant, according to the network state.
It provides several advantages, such as verifiable optimal usage of all devices on the network, better resilience to disconnection, independence from cloud connection, improved privacy and security.

The remainder of this paper is organized in 7 sections.
Section~\ref{sec:motivation} introduces our motivating scenario related to a cultural heritage building and shows the need for a decentralized approach.
Section~\ref{sec:related} overviews relevant related work and highlights the originality of our approach.
Section~\ref{sec:architecture} details our proposed architecture and shows how it drives run-time migration of applications on the edge.
Section~\ref{sec:node_application} presents our network monitoring application and shows how the monitoring takes place.
In Section~\ref{sec:implementation}, we propose a technical implementation, and we validate and evaluate our solution with a proof-of-concept prototype related to our cultural heritage scenario.
Section~\ref{sec:conclusion} discusses the results obtained and gives insights for possible future work.

\section{Motivating Scenario}
\label{sec:motivation}

In this section, we illustrate the relevance of our approach with a scenario related to a Slovenian cultural heritage building located in Bled, Slovenia.
This building has been equipped with multiple sensors to monitor its evolution.
The collected data includes temperature, CO2, relative humidity, Volatile Organic Compounds (VOC), ambient light and atmospheric pressure.
In this scenario, the following constraints motivate the need for a fully decentralized edge computing approach:
\begin{itemize}
    \item Privacy: collected data about the state of the technological solution being deployed is classified as sensitive information.
    Although data about the building could be sent to the cloud, data about the state of resources needs to remain local and only accessible for administration purpose and for the deployed solution to self-manage.
    \item Reliability: centralized orchestration is not appropriate as data collection needs to be resilient to failure of any device. The network of devices needs to adjust to device disconnection any time and keep operating in an optimal way.
    \item Cost: reducing the overall cost by avoiding investing in a cloud infrastructure that involves monthly payments and permanent connection to maintain.
    \item Scalability: as the number of devices will evolve over time, it is necessary for the solution to be able to adjust to changes and homogeneously spread the computation over the network.
    \item Performance: reactivity to external events is improved if processing is performed on-site.
    \item Cost effectiveness: using existing devices that control sensors to perform necessary processing reduces the resource requirements of cloud based solutions, which reduces cost.
\end{itemize}

In this context, it is relevant to equip devices with the capacity to run applications locally and to self-manage the global network load and distribute it over connected devices, according to the state of the network.
In the next section, we present related work and show the need for a decentralized self-managed platform on the edge.
We also overview existing solutions to abstract from platform heterogeneity and justify the technological choice of a container platform to support our solution.

\section{Background Knowledge and Related Work}
\label{sec:related}

\subsection{Orchestration solutions for edge computing}

Strictly observing the definition of orchestration, it always represents control from one party’s perspective.
This differs from choreography, which is more collaborative and allows each involved party to describe its part in the interaction \cite{peltz2003web}. However, to the authors' knowledge, there are no choreography solutions that tackle the problems defined in previous section. Existing orchestration solutions typically rely on a master/slave model where a node is put in charge of the network and decides to allocate applications to nodes according to an optimization algorithm.

Kubernetes~\cite{hightower2017kubernetes} is the most widely used orchestration tool, it is the go-to tool for orchestration in the Google cloud, mostly used in the Microsoft Azure platform and similar products.
It is also the most feature-filled orchestration tool available~\cite{medel2016modelling}.
It has strong community support across many different cloud platforms (in addition to Google cloud, OpenStack, AWS, Azure).

AWS Elastic Container Service (AWS ECS)~\cite{acuna2016amazon}, Amazon’s native container orchestration tool, is the best option for orchestration of AWS services as it is fully integrated into the Amazon ecosystem. It thus integrates easily with other AWS tools.
The biggest limitation is that it is limited to Amazon services.

Docker Swarm~\footnote{\url{https://github.com/docker/swarm}} ships directly with Docker (integrates with Docker-compose) and is supposed to have the simplest configuration.
However, it lacks some advanced monitoring options as compared to other products like Kubernetes.

Apache Mesos’ based DC/OS~\footnote{\url{https://dcos.io/}} is a “distributed operation system” running on private and public cloud infrastructure that abstracts the resources of a cluster of machines and provides common services.

All presented architectures still have a common flaw: single point of failure and a lack of integration with the edge computing.

\subsection{Container platforms}

Containers as used in the purpose of this paper are run as a group of namespaced processes within an operating system, avoiding the overhead of starting and maintaining virtual machines (at the same time providing most of the functionalities). Application containers, such as Docker, encapsulate the files, dependencies and libraries of an application to run on an OS as opposed to the System containers, such as LXC that encapsulate the whole operating system and are in this view more similar to Virtual Machines. The key advantage of containers over virtual machines is their light weight with respect to resources.

Docker~\cite{anderson2015} is the de-facto standard in the open source application container platforms and made containers mainstream.

Core OS’ rkt~\footnote{\url{https://coreos.com/rkt/}}
offers similar functionality as Docker. Rkt is the container runtime from CoreOS.
Like Docker, Rkt is designed for application containers. The market share comparing to Docker is still much lower, but it is raising and with the new announced merges of Redhat and CoreOS in the development, it presents a viable alternative.

LXC~\footnote{\url{https://linuxcontainers.org/}}, short for Linux Containers, is the container runtime and toolset that helped make Docker possible. LXC predates Docker by several years, and Docker was originally based on LXC (it’s not anymore), but LXC gained little traction.

LXD~\footnote{\url{https://linuxcontainers.org/lxd/introduction/}} is a container platform based on LXC. Essentially, LXD provides an API for controlling the LXC library, as well as easy integration into OpenStack. it is backed by Canonical, the company that develops Ubuntu Linux, which is the primary backer of LXD development at the time of writing. 

Unlike Docker and Rkt, LXC and LXD are system containers and as such out of scope of this paper. The selected platform for our research was Docker as it is the most widely used platform and one of the few that can migrate apps at runtime and enables easy communication. The migration is done by pausing the container, dumping the context of the paused container, transferring the context on a different host that can resume the execution given the context.

\subsection{Decentralized Self-managing IoT Architectures}

A lot of work have proposed solutions to enable fully decentralized self-managing architectures for the IoT.
For example, in~\cite{maior2014self}, the work focuses on a decentralized solution for energy management in IoT architectures connected to smart power grids.
In~\cite{higgins2011distributed}, the authors propose a distributed IoT approach for electrical power demand management problems based on “distributed intelligence” rather than “traditional centralized control,” with the system improving on many levels. Then, in~\cite{suzdalenko2013instantaneous} the authors further develop the former approach by creating a decentralized distributed model of an IoT; where consumers can freely join and leave the system automatically at any time. In~\cite{niyato2011machine} a system that uses machine-to-machine (M2M) communication is presented, to reduce the costs of a home energy management system. Also, dSUMO~\cite{bragard2017self}, a distributed and decentralized microscopic simulation that eliminates the central entity and thus overcome the bottleneck in synchronization. In~\cite{al2018energy}, the authors demonstrate the effectiveness of utilizing a publish/subscribe messaging model as connection means for indoor localization utilizing Wireless Sensor Networks (WSNs) through a middle-ware, the results showed that RSS get an acceptable accuracy for multiple types of applications.

However, all the aforementioned contributions are different from the solution we propose in this paper, at two levels.
First, they mostly focus on a single specific aspect and find an optimal solution for it, without considering the fact that an IoT architecture involves multiple criteria that require optimization.
In our work, we already consider multiple criteria to optimize application migration, while envisioning that this number of criteria can increase in the future.
Second, as far as we know, there is no approach that combines blockchain-like data structure and consensus algorithms in a single framework with the objective to drive application migration at run-time on the edge, which is the main contribution of this paper.

\section{A Decentralized Self-managing Architecture}
\label{sec:architecture}


In the following, we describe the general architecture that support our edge computing platform. Devices on the edge are nodes running node software and a containerization software. A node can join the network by following a network protocol for exchanging known nodes and participating by executing the consensus algorithm. Nodes keep discovering the network by asking connected nodes for peers.
For the sake of simplicity, in this paper we consider that the number of nodes remains reasonably limited, so that large scale discovery issues remain out of the scope of this paper.


Our devices are equipped to allow a specific containerized application (called node app) to introspect the state of the node and handle the diffusion of this information over the network.
It also is responsible for maintaining the information about the other nodes up to date, for participating in the consensus algorithm, and for listening to messages coming from the exposed node API.

\begin{figure}[ht!]
\centering
\includegraphics[width=0.7\linewidth]{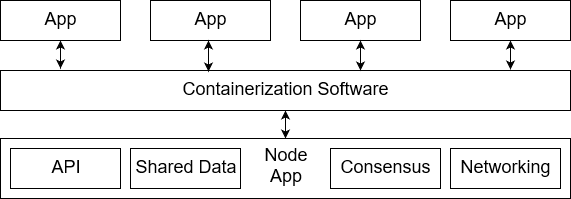}
\caption[Node]{Architecture of an edge device software platform}
\label{fig:node}
\end{figure}

Figure~\ref{fig:node} shows the key components of Nodes in the system. The node software is compiled into a container, in our case Docker. The container mounts a direct socket to the containerization service for querying the state of the system and managing local containers.

\section{Node Application}
\label{sec:node_application}

Every 500 milliseconds, each device collects information about the state of its neighbours.
Typically, a state is a vector of scores that describes the device state and the applications being executed by the node. 
In this work we define a state to be a matrix of vectors $$S (APP,CPU,RAM,DISK,NETWORK,TIMESTAMP)$$ where each vector represents an application being executed by the node and the corresponding resource consumption.
Resources are reported as a fraction of the total available. In order to have comparable values between nodes, reporting on CPU usage and network utilization require some engineering which is outside of the scope of this paper.


Monitoring resources within the P2P network is done by having nodes maintain a list of scores of other nodes. All nodes periodically broadcast digitally signed messages containing their score. All nodes follow simple P2P broadcasting rules that guarantee finality and efficiency in message propagation.
\begin{itemize}
    \item If elapsed time greater then $\Delta ST$, broadcast signed message containing own score.
    \item When receiving a new score message, check if message was received before (compare digital signatures)
    \item If message was not seen before, broadcast it to all connected nodes with the exception of originating node
\end{itemize}
Where $\Delta ST$ is configurable and should depend on the time interval of the consensus algorithm. The score pool hence contains scores of all nodes participating in the network. Each score has a corresponding time-stamp which is later used by elected nodes to create a migration strategy.

For improved efficiency, every score message broadcast is prefaced with a "Do you need this" (DYNT) message coupled with the digital signature of the message only. Messages are sent to nodes that reply to the DYNT message to minimize bandwidth use.

\subsection{Consensus algorithm}
\label{subsec:consensus}
The network requires a consensus algorithm to avoid race conditions when migrating applications.
The choice of a consensus algorithm depends on the requirements of the implementation and domain of application.
In general, any consensus based on leader election can be plugged in.
Examples of such consensus algorithms are Paxos~\cite{lamport2001paxos}, Raft~\cite{ongaro2014search}, PoET~\cite{olson2018sawtooth}, etc.

The elected leader is responsible for creating a migration plan and including the resource consumption estimates in a block. The block gets digitally signed so other nodes can verify it originates from the elected leader.
Nodes receiving a new block must verify the migration plan by computing it locally and comparing the results. If the migration plan is equal, they act on it, otherwise discard the block and wait for a new one. With these simple protocol rules in place the network is Byzantine fault tolerant \cite{castro1999practical}. 

A migration strategy is analogous to a block in block-chain based systems. Blocks contain all the data shared among nodes in the network and include a digital signature of the previous block thus creating a block chain. In order to create a digital signature of block $n +1 $ a node needs to have the digital signature of node $n$. A well formed block can be verified by other nodes that also have block $n$. In case of a malformed block, verification will fail, and nodes will reject the block, thus forcing the nodes to agree on the shared data.
The block serves as an instruction set mapping applications to nodes.
Consider a case with 4 nodes in set $N$ denoted by $A, B, C$, and $D$ respectively.
All nodes share their score and keep a local copy of reported scores of other nodes.
Each node also stores a vector of applications $v \in V$ that need to be executed.
Table~\ref{table:block} shows an example of a block $k$ which assigns every $v \in V$ to a node $n \in N$ To create block $k+1$ a node elected as leader computes an assignment such that the use of resources is optimal.
The input to the algorithm is limited to block data to ensure determinism that can enforce consensus.
The algorithm depends on the application domain and exploring available possibilities will be subject to future work.
In this paper, we use the simple algorithm described below, which is deterministic and can only take the block data as input for computation.


\begin{algorithm}[ht!]
\SetAlgoLined
\KwData{BlockData}
\KwResult{Migration plan}
$Max \gets FindMaxLoadedNode(BlockData)$\;
$Min \gets FindMinLoadedNode(BlockData)$\;

\eIf{!AppQueue.isEmpty()}{
 \While{!AppQueue.isEmpty()}{
 $Min \gets  FindMinLoadedNode(BlockData)$\;
 $Min.addApp(AppQueue.dequeue())$\;
 }
 }{
  $AppToMigrate \gets  Max.MaxLoadApp$\;
  $CurrentDeltaScore \gets  (Max.score -Min.score)$\;
  $FutureDeltaScore \gets (Max.score - AppToMigrate.score ) - (Min.score + AppToMigrate.score)$\;
  
  \If{$Math.abs(CurrentDeltaScore > FutureDeltaScore$)}
  {
    Migrate $AppToMigrate$ to $Min$\;
  }
 }
 \caption{Deterministic migration plan generation algorithm}
\end{algorithm}

\begin{center}
\begin{table}[ht]
\caption{Block data} 
\centering 
\begin{tabular}{c c c c c c} 
\hline 
V&Node&RAM&DISK&CPU&Average Latency \\ [0.5ex] 
\hline 
$v_0$ & A & 50\% & 23\% & 90\% & 23ms \\ 
$v_1$ & B & 47\% & 87\% & 23\% & 33ms\\
$v_2$ & C & 12\% & 25\% & 15\% & 51ms \\
$v_3$ & A & 35\% & 14\% & 56\% & 101ms \\
$v_4$ & D & 25\% & 74\% & 16\% & 9ms \\[1ex] 
\hline 
\end{tabular}
\label{table:block} 
\end{table}
\end{center}

Once a block is created, currently reported scores are included that will be used to compute block $k + 2$. Additionally, blocks are equipped with meta-data like block hash, previous block hash, etc. to facilitate their utilization.


\section{Implementation and Evaluation}
\label{sec:implementation}

\subsection{Technical Implementation}

As described in Section~\ref{sec:motivation}, we have implemented and evaluated our solution with a set of sensors deployed in the cultural heritage building Mrakova Domačija in Bled, Slovenia.
Each sensor is connected to a Raspberry Pi device that hosts a Linux Alpine OS and a Docker container.
We developed our node application inside a container, it relies on the Docker introspection capacity (\texttt{docker stats} command called from our Java program) to collect information about each device.
The application also hosts a HTTP server\footnote{Please note that CoAP could be used for energy saving purposes.} that allows communicating with other nodes through a RESTful API operating as follows:

\begin{itemize}
    \item HTTP GET gives a representation of the target node, which includes information about the state of the device as well as all the necessary information about the node (i.e. last connection time, average connection time\dots).
    \item HTTP PUT sends information to the target node about the state of the source node.
    Such request is useful for nodes to send to their neighbours information about their current state.
    HTTP PUT allows system designers to specify URLs where shared information is stored (for example~\url{http://192.168.1.15/shared}).
    \item HTTP POST holds the same role as HTTP PUT but it applies to new devices, so that the data is added to the shared pool and does not replace existing data.
    \item HTTP DELETE is utilized when a node leaves the network in a predictable way, so that its state information is removed from the shared pool without going through a time-out.
\end{itemize}

\subsection{Validation and Evaluation}

To validate the feasibility of our approach and test its scalability we ran performance simulation test cases. In each test case, a fixed number of nodes formed a P2P network.
Nodes were assigned applications to execute.
Each application had a random execution time and preset resource consumption expressed in fractions between 5\% - 40\%. For the sake of simplicity, only one resource was used (CPU). 
The simulation ran for 100 blocks with a block time of 1 second.
Applications were queued until the average load of the entire system rose above 90\%. The migration strategy was implemented based on the algorithm described in Section~\ref{subsec:consensus}. Applications arrived in the queue with certain probability, which was gradually increased with the number of nodes in the system.
From the reported resource loads of nodes (reported in \%), we compute the standard deviation as a measure of how balanced resource consumption is. 

\begin{figure}[ht!]
\centering
\begin{subfigure}{.5\textwidth}
  \centering
  \includegraphics[width=\linewidth]{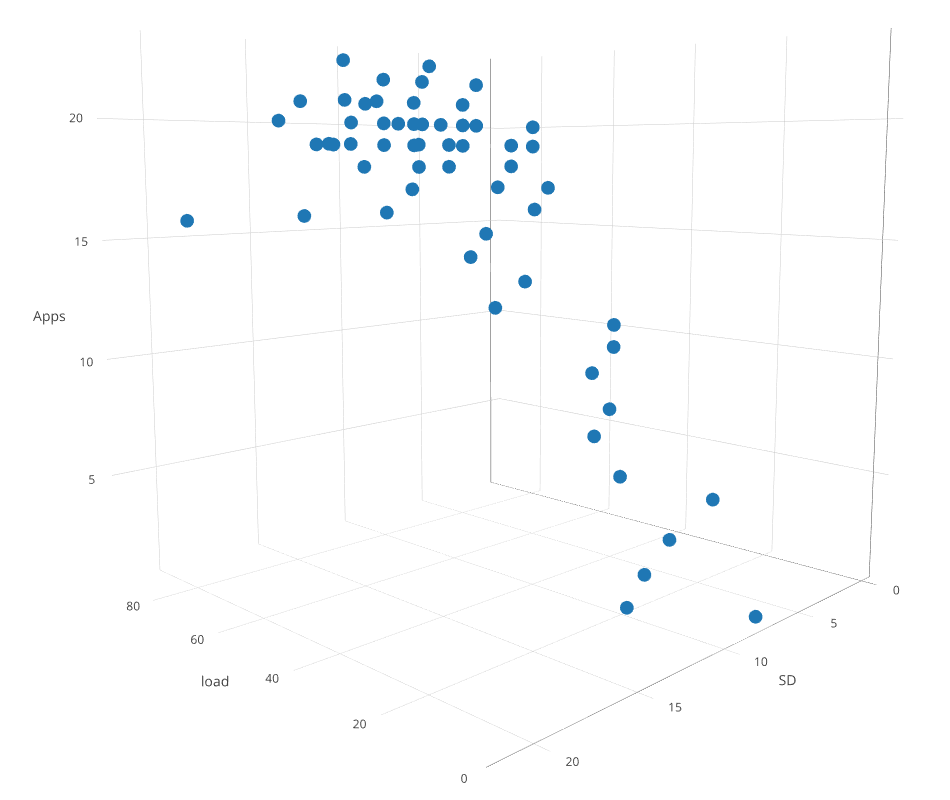}
\caption[Node]{5 nodes}
\label{fig:5nodes}
\end{subfigure}%
\begin{subfigure}{.5\textwidth}
  \centering
    \includegraphics[width=\linewidth]{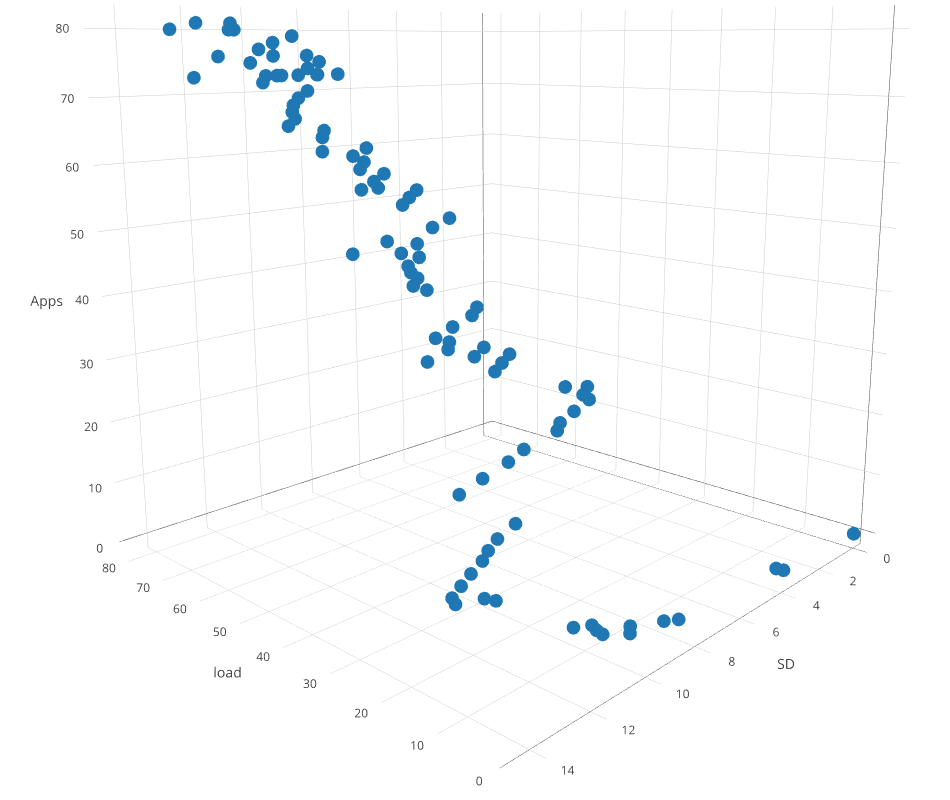}
    \caption[Node]{25 nodes}
    \label{fig:25nodes}
\end{subfigure}

\medskip
\begin{subfigure}{.5\textwidth}
  \centering
    \includegraphics[width=\linewidth]{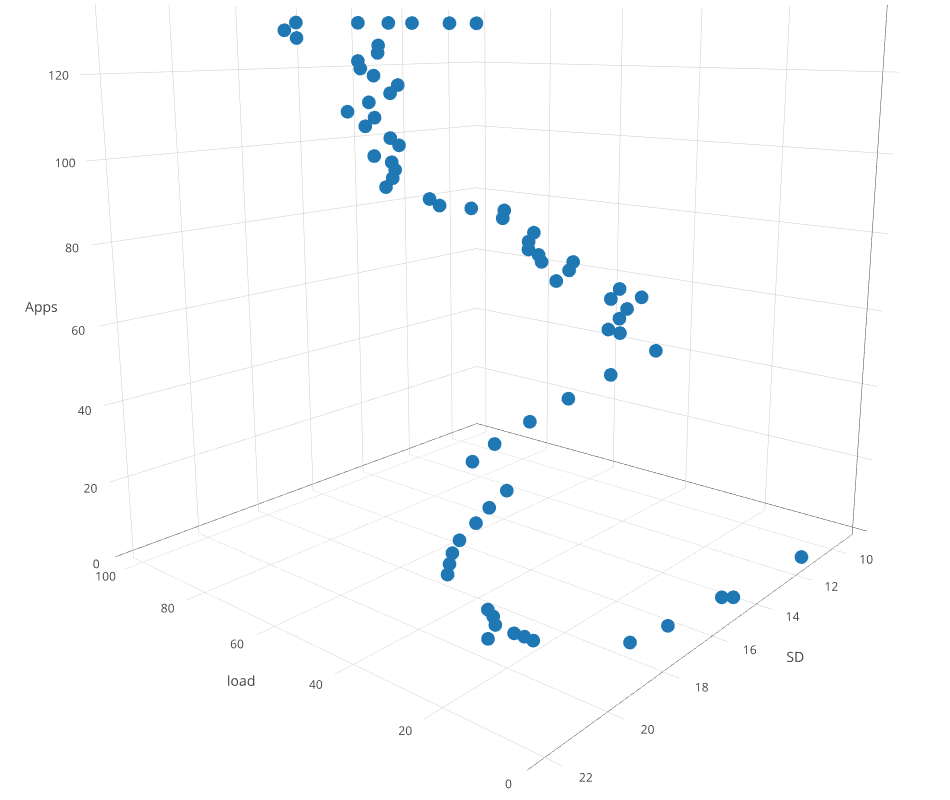}
    \caption[Node]{50 nodes}
    \label{fig:50nodes}
\end{subfigure}%
\begin{subfigure}{.5\textwidth}
  \centering
    \includegraphics[width=\linewidth]{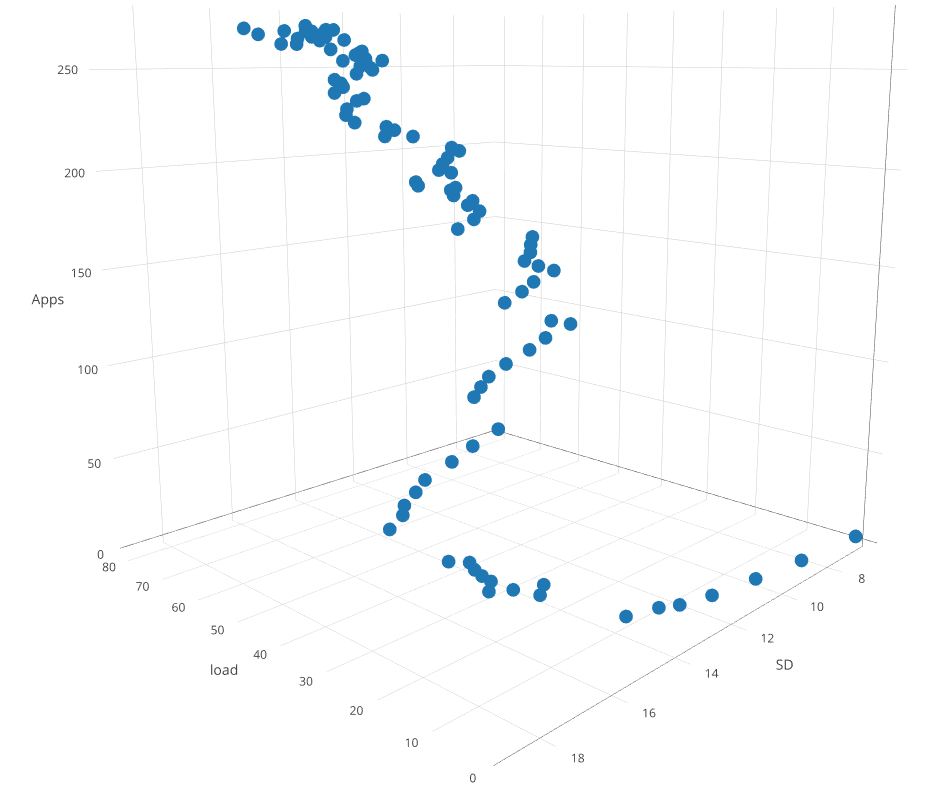}
    \caption[Node]{100 nodes}
    \label{fig:100nodes}
\end{subfigure}
\caption{Simulation results}
\label{fig:test}
\end{figure}

In Fig.~\ref{fig:test}, we observe that the standard deviation remains low even when the number of applications in the system grows.
The lower load cases where we can observe higher swings in standard deviations are expected due to the low number of applications.
The crossover happens when the number of applications exceeds the number of nodes.
Below the threshold, there are bound to be nodes that do not run any applications.
We can observe from Fig.~\ref{fig:5nodes} that as the number of nodes is low, resource balancing between nodes is effective earlier, which explains why the measures are less marked than with the other figures, that correspond to test cases where it takes the simulation a longer time to reach the point of crossover where a higher number of applications is distributed over a lower number of nodes.

From the simulation results we conclude that the architecture can scale with the growing number of nodes in the network. Additionally, the naive algorithm for creating a migration strategy performed well in distributing load across the system.

\section{Discussion and Conclusion}
\label{sec:conclusion}

In this paper, we propose a decentralized solution to the resource usage optimization problem, a typical issue in edge computing.
Our solution avoids the single point of failure that centralized architectures suffer from and improves network resilience as it does not depend on a master node.
To design our solution, we have combined a blockchain-like shared data structure and a consensus algorithm with a monitoring application that runs on top of the Docker platform.
Such combination allows edge devices to check at run-time if there is a need for migrating an application, and to reach consensus on a decision to do so.
With our contribution, edge devices become a completely decentralized and distributed run-time platform.
We have implemented and evaluated our solution with a set of sensors deployed in a cultural heritage building in Bled, Slovenia.

Results show that our approach is able to adjust and normalize the application load over a set of nodes.
It also provides, thanks to the fact that the algorithm we use is deterministic and that all the data is stored in a distributed structure, the possibility to verify all the decisions that have been taken to optimize the usage of edge devices.
The consensus algorithm that we use also allows to adjust the global network behaviour to entering or leaving nodes.

Several limitations have been identified that give insights for future work.
First, it is important to observe how adding and removing devices affects network behaviour and to explore how scalable is our approach over a large number of devices.
Second, it seems appropriate to find out what specific aspects of use cases can help determine which consensus algorithm is most suitable for deploying our solution, in order to best match the use case requirements.
Third, it includes semantically describing applications and the services that edge devices offer, to support application migration, and combine in the same architecture the need for efficiently managing network resources together with the needs of applications in terms of functionality and quality of service.

 \section{Acknowledgment}
\label{acknowledgment}
The authors gratefully acknowledge the European Commission for funding the InnoRenew CoE project (Grant Agreement \#739574) under the Horizon2020 Widespread-Teaming program and and the Republic of Slovenia (Investment funding of the Republic of Slovenia and the European Union of the European regional Development Fund).

%
%
%
\bibliographystyle{splncs04}
\bibliography{ref}

\end{document}